\documentclass[onecolumn,10pt]{article}
\usepackage[top=2.54cm, bottom=2.54cm, left=2.54cm, right=2.54cm,a4paper]{geometry}

\usepackage{amsmath,amsthm}
\usepackage{bm}  
\usepackage{bbm}
\usepackage{verbatim}
\usepackage{xcolor}

\usepackage{qcircuit}

\usepackage[backend=biber,sorting=none,style=phys,biblabel=brackets,backref=false,bibencoding=utf8,maxnames=20,eprint=true]{biblatex}
\makeatletter
\pretocmd{\blx@head@bibintoc}{\phantomsection}{}{\ddt}
\makeatother
\DefineBibliographyStrings{english}{%
  backrefpage = {page},
  backrefpages = {pages},
}
\usepackage{titlesec}

\titleformat*{\section}{\bfseries}
\titleformat*{\subsection}{\normalsize\bfseries}
\titleformat*{\subsubsection}{\bfseries}
\titleformat*{\paragraph}{\large\bfseries}
\titleformat*{\subparagraph}{\large\bfseries}

\titlespacing\section{0pt}{12pt plus 4pt minus 2pt}{2pt plus 2pt minus 2pt}

\usepackage[bookmarks,colorlinks,breaklinks]{hyperref}  
\definecolor{dullmagenta}{rgb}{0.4,0,0.4}   
\definecolor{darkblue}{rgb}{0,0,0.4}
\hypersetup{linkcolor=red,citecolor=blue,filecolor=dullmagenta,urlcolor=darkblue} 

\usepackage[bitstream-charter,cal=cmcal]{mathdesign}
\usepackage[T1]{fontenc}

\usepackage{amsbsy,verbatim,graphicx}
\usepackage{units}

\newcommand{\ket}[1]{|#1\rangle}

\newcommand{\bra}[1]{\langle #1|}

\newcommand{\ketbra}[1]{\ket{#1}\bra{#1}}

\newcommand{\tr}{{\text{Tr}}}
\newcommand{\id}{\mathbbm 1}

\def\cE{\mathcal E}

\def\cI{\mathcal I}

\def\cM{\mathcal M}

\def\cT{\mathcal T}

\def\xsize{{|X|}}

\newtheorem{prop}{Proposition}

\usepackage{xfrac}

\usepackage{setspace}
\usepackage{titling}

\posttitle{\par\end{center}}
\predate{}
\postdate{}
\begin{document}
\author{
{\normalsize Joseph M.\ Renes}\\
\emph{\normalsize  
Institute for Theoretical Physics, ETH Z\"urich, Switzerland}
}



\title{\large {\bf Better bounds on optimal measurement and entanglement recovery, with applications to uncertainty and monogamy relations}}

\date{\vspace{-\baselineskip}}

\maketitle

\begin{abstract}
We extend the recent bounds of Sason and Verd\'u relating R\'enyi entropy and Bayesian hypothesis testing [\href{http://arxiv.org/abs/1701.01974}{arXiv:1701.01974}] to the quantum domain and show that they have a number of different applications. 
First, we obtain a sharper bound relating the optimal probability of correctly distinguishing elements of an ensemble of states to that of the pretty good measurement, and an analogous bound for optimal and pretty good entanglement recovery. 
Second, we obtain bounds relating optimal guessing and entanglement recovery to the fidelity of the state with a product state, which then leads to tight tripartite uncertainty and monogamy relations. 
\end{abstract}
\vspace{0.5\baselineskip}

\section{Introduction}

Successful analysis of information processing protocols requires suitable measures of information and entropy, particularly those that satisfy the data processing inequality, the statement that a formal measure of information satisfies the intuitive requirement that a noisy channel cannot increase it. 
One broad class of measures is given by the R\'enyi divergences, which includes the usual Shannon and von Neumann definitions of mutual information and entropy. 
But even more, the R\'enyi divergences also encompass optimal and ``pretty good'' strategies for distinguishing quantum states or recovering entanglement, and are related to the oft-used fidelity function. 
Hence new insights into these measures often leads to new results for these operational tasks. 
This is the case in \cite{iten_pretty_2017}, for instance, which found new conditions for the optimality of the pretty good measurement by investigating the relationship of various quantum R\'enyi divergences.

Here we extend a recent result by Sason and Verd\'u~\cite{sason_arimoto-renyi_2017}, which establishes a whole class of Fano-like inequalities involving the R\'enyi divergence and optimal distinguishing probability, to the quantum domain. 
Though the inequalities are essentially an immediate consequence of the data processing inequality, they turn out to have a number of interesting applications. 
First, we find improved bounds relating the pretty good measurement to the optimal measurement, as well as analogous bounds for pretty good and optimal entanglement recovery.
Second, by establishing a new relation between the optimal guessing probability and the fidelity, we can provide a complete characterization of the set of admissible guessing probabilities in an uncertainty game~\cite{berta_uncertainty_2010,renes_uncertainty_2016-1}, which is also related to wave-particle duality relations in multiport interferometers~\cite{coles_entropic_2016}. 
The goal of game is to provide predictions of the values of potential measurements of two conjugate observables on a quantum system; the uncertainty principle implies that the predictions cannot both be accurate. 
The same relation holds for fidelity and optimal entanglement recovery, and in this context gives a complete characterization of the possible entanglement fidelities two  different parties can have with a common system, resolving a conjecture for the ``singlet monogamy'' studied in~\cite{kay_optimal_2009}.

\section{Setup}
\subsection{R\'enyi divergences}
For two classical distributions $P$ and $Q$, the R\'enyi divergence of order $\alpha\in (0,1)\cup(1,\infty)$ is given by~\cite{renyi_measures_1961} 
\begin{align}
D_\alpha(P,Q):=\frac1{\alpha-1}\log \sum_x P(x)^\alpha Q(x)^{1-\alpha}\,.
\end{align}
The limits $\alpha=0,1,\infty$ are obtained by continuity in $\alpha$, which gives $D_0(P,Q)=-\log \sum_{x:P(x)>0}Q(x)$, $D_1(P,Q)=\sum_x P(x)\log \frac{P(x)}{Q(x)}$, and $D_\infty(P,Q)=\max_{x}\log \frac{P(x)}{Q(x)}$. 
In the case of distributions on a binary-valued random variable it will be convenient to define the binary R\'enyi divergence
\begin{align}
d_\alpha(p,q):=\tfrac1{\alpha-1}\log\left(p^\alpha q^{1-\alpha}+(1-p)^\alpha(1-q)^{1-\alpha}\right)\,,
\end{align}
where $0\leq p\leq 1$ and $0\leq q\leq 1$. Again the limiting cases are established by continuity.

In the quantum case there are several possible definitions of the R\'enyi divergence, due to the choices of ordering the density operators. 
Only those that satisfy the data processing inequality are useful for our purposes, and there are several; see \cite{tomamichel_quantum_2016} for an overview.
Here we will focus on the minimal, or sandwiched, R\'enyi divergence~\cite{muller-lennert_quantum_2013,wilde_strong_2014},
\begin{align}
D_\alpha(\rho,\sigma)&:=\frac1{\alpha-1}\log \tr\left[\left(\sigma^{\frac{1-\alpha}{2\alpha}}\rho\sigma^{\frac{1-\alpha}{2\alpha}}\right)^\alpha\right]\,.
\end{align}
The name ``minimal'' comes from the fact that quantity is the smallest in a family of possible R\'enyi divergences which satisfies the data processing inequality (in this case, for $\alpha\in [\tfrac12,\infty]$)~\cite{tomamichel_quantum_2016}.

Though we will not explicitly make use of it here, we mention that the R\'enyi divergence can be used to define a conditional entropy by either $H_\alpha^\downarrow (X|B)_\rho:=\log |X|-D_\alpha(\rho_{XB},\pi_X\otimes \rho_B)$ or $H^\uparrow_\alpha(X|B)_\rho:=\sup_\sigma(\log |X|-D_\alpha(\rho_{XB},\pi_X\otimes \sigma_B)$. 
In the classical case this definition goes back to Arimoto~\cite{arimoto_information_1977} and appears to be the most meaningful extension of the usual Shannon conditional entropy to the R\'enyi setting~\cite{teixeira_conditional_2012,fehr_conditional_2014}. 

\subsection{Guessing probabilities and entanglement recovery}
The minimal divergence is also interesting due to its connections with the fidelity function as well as optimal and ``pretty good'' guessing probabilities and entanglement recovery. 
For $F(\rho,\sigma)=\|\sqrt\rho\sqrt\sigma\|_1$ the fidelity of the states $\rho$ and $\sigma$, we have $D_{\nicefrac12}(\rho,\sigma)=-\log F(\rho,\sigma)^2$. 

An arbitrary ensemble of states $\varphi_x$ with prior probabilities $p_x$ can be encapsulated in the classical-quantum state $\rho_{XB}=\sum_x p_x \ketbra x_x\otimes (\varphi_x)_B$. 
Any given measurement $\Lambda$ on $B$ results in some average probability of correctly guessing, $P(X|B)_{\rho,\Lambda}=\sum_x p_x \tr[\varphi_x \Lambda_x]$. 
As shown in \cite{konig_operational_2009}, the optimal probability $P_\text{opt}$ satisfies
\begin{align}
\label{eq:optm}
\inf_\sigma D_\infty(\rho_{XB},\pi_X\otimes \sigma_B)=\log \xsize P_\text{opt}(X|B)_\rho\,,
\end{align} 
where $\pi_X$ is the completely mixed state (uniform distribution). 
The ``pretty good measurement''~\cite{belavkin_optimal_1975,hausladen_`pretty_1994} uses the POVM elements $\Lambda_x=\varphi^{-\nicefrac12}p_x\varphi_x \varphi^{-\nicefrac12}$ for $\varphi=\sum_x p_x\varphi_x$, and its guessing probability $P_\text{pg}$ satisfies~\cite{buhrman_possibility_2008}
\begin{align}
\label{eq:pgm}
 D_2(\rho_{XB},\pi_X\otimes \rho_B)=\log \xsize P_\text{pg}(X|B)_\rho\,.
\end{align}
When the $\varphi_x$ commute and $B$ is effectively a classical random variable $Y$, the pretty good measurement reduces to guessing $X$ by sampling from the distribution $P_{X|Y=y}$ for the observed value of $Y$. Beyond its use in quantum information theory, this measurement has also been used to construct decoders for error-correcting codes in classical information theory~\cite{yassaee_technique_2013-1}. 

The fully quantum analog of the guessing scenario is that of entanglement recovery by local action. 
For an arbitrary bipartite entangled state $\rho_{AB}$. a quantum channel $\cE_{A'|B}$ taking $B$ to $A'\simeq A$ results in some (squared) fidelity with the maximally entangled state $\ket{\Phi}_{AA'}=\tfrac1{\sqrt{|A|}}\sum_x \ket x_A\ket x_{A'}$, $R(A|B)_{\rho,\cE}=\tr[\Phi_{AA'}\cE_{A'|B}(\rho_{AB})]$. 
The optimal fidelity $R_\text{opt}$ satisfies~\cite{konig_operational_2009} 
\begin{align}
\label{eq:optr}
\inf_\sigma D_\infty(\rho_{AB},\pi_A\otimes \sigma_B)=\log |A|^2 R_\text{opt}(A|B)_\rho\,.
\end{align}
The ``pretty good recovery'' uses the map $\cE_{A'|B}(\sigma_{AB})=\tr_B[\rho_B^{-\nicefrac12}\rho_{A'B}\rho_B^{-\nicefrac12}\sigma_{AB}^{T_B}]$ and satisfies~\cite{berta_entanglement-assisted_2014}  
\begin{align}
\label{eq:pgr}
D_2(\rho_{AB},\pi_A\otimes \rho_B)=\log |A|^2 R_\text{pg}(A|B)_\rho\,.
\end{align}
When $\tr_B[\rho_{AB}]=\pi_A$, this is the recovery map of \cite{barnum_reversing_2002}. 

\section{Bounds}

Regarding the POVM $\{\Lambda_x\}$ as the quantum-classical channel $\cM_{X'|B}$, we can express the guessing probability as $P(X|B)_{\rho,\Lambda}=\tr[\Pi_{XX'}\cM_{X'|B}(\rho_{XB})]$,
where $\Pi_{XX'}=\sum_{x\in \mathcal X}\ketbra x_X\otimes \ketbra x_{X'}$.  
Note that $\tr[\Pi_{XX'}\cM_{X'|B}(\pi_X\otimes \sigma_B)]=\frac1{\xsize}$ for any state $\sigma_B$. 
The projector $\Pi_{XX'}$ is part of a two-outcome measurement, a test, described by the channel $\cT_{Y|XX'}$. 
The random variable $Y$ equals 1 when the test passes, corresponding to $\Pi_{XX'}$, and zero if it fails, corresponding to $\id_{XX'}-\Pi_{XX'}$.
Similarly, the expression for $R(A|B)_{\rho,\cE}$ makes use of the test $\cT_{Y|AA'}$  involving $\Phi_{AA'}$. 
And in this case we have, for any $\sigma_{A'}$,  $\tr[\Phi_{AA'}\pi_A\otimes \sigma_{A'}]=\tfrac1{|A|^2}$.
Applying the data processing inequality of the R\'enyi divergence for $\cT\circ \cM$ or $\cT\circ \cE$ immediately gives our main result. 
\begin{prop}
Let $\sigma_B$ be any normalized state and $\alpha\in [\tfrac12,\infty]$. 
For arbitrary classical-quantum states $\rho_{XB}$ and measurements $\{\Lambda_x\}$ on $B$, we have
\begin{align}
\label{eq:mainc}
 D_\alpha(\rho_{XB},\pi_X\otimes \sigma_B)\geq d_\alpha(P(X|B)_{\rho,\Lambda},\tfrac1{\xsize})\,.
\end{align}
For $\rho_{AB}$ an arbitrary bipartite quantum state and $\cE_{A'|B}$ a quantum channel from $B$ to $A'\simeq A$, we have
\begin{align}
\label{eq:mainq}
D_\alpha(\rho_{AB},\pi_A\otimes \sigma_B)\geq d_\alpha(R(A|B)_{\rho,\cE},\tfrac1{|A|^2})\,.
\end{align}
\end{prop}

Choosing $\alpha=1$, $\sigma_B=\rho_B$, and the optimal measurement $\Lambda$ or recovery map $\cE_{A'|B}$ gives the Fano inequalities
\begin{align}
H(X|B)_\rho&\leq (1-P_\text{opt}(X|B)_\rho)\log (|X|-1)+h_2(P_\text{opt}(X|B)_\rho)\,, \qquad \text{and}\\
H(A|B)_\rho&\leq -\log |A|+(1-R_\text{opt}(A|B)_\rho)\log(|A|^2-1)+h_2(R_\text{opt}(A|B)_\rho)\,,
\end{align}
where $h_2(x)=-x \log_2 x -(1-x)\log_2(1-x)$ is the binary entropy.

\subsection{Pretty good measurement and entanglement recovery}
Choosing $\cM_{X'|Y}$ to be the optimal measurement for given $\rho_{XB}$ and using \eqref{eq:pgm} in \eqref{eq:mainc} gives 
\begin{align}
\label{eq:pgmbound}
P_\text{pg}(X|B)_\rho\geq {P_\text{opt}(X|B)_\rho^2}+\frac{(1-P_\text{opt}(X|B)_\rho)^2}{\xsize-1}\,.
\end{align}
In the classical case this was first shown by \cite[Theorem 3]{devijver_new_1974}, though without the connection between $D_2$ and $P_\text{pg}(X|B)_\rho$.

Equality can be attained (also shown in \cite{devijver_new_1974}), as illustrated by the ``L distribution'' with weight $p_0$ on $X=0$ and $(1-p_0)/(\xsize-1)$. 
The optimal guess is always $X=0$, meaning $P_\text{opt}(X)=p_0$. 
Meanwhile, the pretty good measurement generates its guess in this case by sampling from the distribution. 
Therefore $P_\text{pg}(X)=p_0^2+(\xsize-1)\left(\frac{1-p_0}{\xsize-1}\right)^2$, which is precisely the righthand side above. 

Taking the limit $\xsize\to \infty$, we recover the previously-known result, $P_\text{pg}(X|B)_\rho\geq P_\text{opt}(X|B)_\rho^2$, first shown in~\cite{vajda_bounds_1968} in the classical case (again, just as a statement involving on $D_2$) and in~\cite{barnum_reversing_2002} in the quantum case. 
The new bound resolves a defect of the previous bound, in that the value of the new bound is always larger than $1/\xsize$. 
To see this, observe that the righthand side of \eqref{eq:pgmbound} minus $1/\xsize$ is simply $\tfrac{(\xsize P_\text{opt})^2}{\xsize(\xsize-1)}>0$.
This ensures that the bound is meaningful for any value of $P_\text{opt}(X|B)_\rho$, whereas the previous bound is only meaningful when $P_\text{opt}(X|B)_\rho\geq 1/\sqrt{|X|}$.

Choosing $\cE_{A'|B}$ to be the optimal recovery map for given $\rho_{AB}$ and using \eqref{eq:pgr} in \eqref{eq:mainq} similarly gives 
\begin{align}
R_\text{pg}(A|B)_\rho\geq {R_\text{opt}(A|B)_\rho^2}+\frac{(1-R_\text{opt}(A|B)_\rho)^2}{|A|^2-1}\,.
\end{align}
Equality can also be attained in this bound, by essentially the same example. 
Suppose $\rho_{AB}$ is a Bell-diagonal state with weight $p_0$ on $\ket{\Phi}$ and $1-p_0$ evenly spread over the remaining $|A|^2-1$ Bell states. The local state on system $B$ is the same for all Bell states, so there is no advantage to applying a nontrivial recovery map on $B$; hence $R_\text{opt}(A|B)_\rho=p_0$. 
On the other hand, using the pretty good recovery leads to $R_\text{pg}(A|B)_\rho=\tr[\rho_{AB}^2]$, which then gives the righthand side.

Bounds in the other direction can be obtained by choosing $\cM$ to be the pretty good measurement and using \eqref{eq:optm} in \eqref{eq:mainc}, or $\cE$ to be the pretty good recovery and using \eqref{eq:optr} in \eqref{eq:mainq}.   
However, this leads back to the obvious lower bounds $P_\text{opt}(X|B)_\rho\geq P_\text{pg}(X|B)_\rho$ and $R_\text{opt}(A|B)_\rho\geq R_\text{pg}(A|B)_\rho$.

\subsection{Uncertainty and monogamy relations}
Again choosing the optimal measurement or recovery map but now using the relationship between $D_{\nicefrac 12}$ and the fidelity gives
\begin{align}
\label{eq:maxbound}
F(\rho_{XB},\pi_X\otimes \sigma_B)^2
&\leq  \frac1{\xsize}\left(\sqrt{P_\text{opt}(X|B)_\rho}+\sqrt{\xsize-1}\sqrt{1-P_\text{opt}(X|B)_\rho}\,\right)^2\,,\\
\label{eq:maxboundq}
F(\rho_{AB},\pi_A\otimes \sigma_B)^2
&\leq \frac1{|A|^2}\left(\sqrt{R_\text{opt}(A|B)_\rho}+\sqrt{|A|^2-1}\sqrt{1-R_\text{opt}(A|B)_\rho}\,\right)^2\,.
\end{align}
In the case of classical $B$ the former bound was reported by Sason and Verd\'u~\cite[Equation~109]{sason_arimoto-renyi_2017}. 
Employing the ``L distribution'' again yields equality in both. 
Thus, the former is necessarily stronger than the bound reported by the author in~\cite[Equation~23]{renes_uncertainty_2016-1} 
as well the bound discovered by Coles~\cite[Equation~6]{coles_entropic_2016}, 

We can use \eqref{eq:maxbound} to completely characterize the region of allowed guessing probabilities in the three party uncertainty game considered in~\cite{renes_uncertainty_2016-1}. 
Suppose $\rho_{ABC}$ is a tripartite quantum state and $\psi_{XB}$ is the classical-quantum state resulting from measuring an observable $X$ on system $A$ and ignoring $C$, while $\xi_{ZC}$ is the classical-quantum state resulting from measuring the conjugate observable $Z$ on $A$ and ignoring $B$. 
An immediate question is what are the allowed values of $P(X|B)_{\psi,\Lambda}$ and $P(Z|C)_{\xi,\Gamma}$. 
To determine the boundary of the set, start with the uncertainty relation for min and max entropy~\cite{tomamichel_uncertainty_2011}, which can be expressed as 
$\max_\sigma F(\psi_{XB},\pi_X\otimes \sigma_B)^2\geq P_\text{opt}(Z|C)_\xi$.
Combining this with \eqref{eq:maxbound} gives
\begin{align}
\label{eq:tripartiteboundary}
|A| P_\text{opt}(Z|C)_\xi\leq \left( \sqrt{P_\text{opt}(X|B)_\psi}+\sqrt{|A|-1}\sqrt{1-P_\text{opt}(X|B)_\psi}\,  \right)^2\,.
\end{align}
In principle, we could also interchange the two guessing probabilities to obtain another bound, but in fact this leads back to the same inequality. 
The bound also tightens the relation between fringe visibility and path distinguishability in symmetric multipath interferometers, Theorem 1 of \cite{coles_entropic_2016}, as these quantities are rescaled versions of the two guessing probabilities. 

Equality can be attained in \eqref{eq:tripartiteboundary} by, unsurprisingly, a state involving an ``L distribution''. 
In particular, consider the case of trivial $B$ and $C$, and $\ket{\theta}_A$ the state with amplitudes $\sqrt{p_0}$ for $\ket 0$ and $\sqrt{\frac{1-p_0}{|A|-1}}$ for $\ket x$ with $x\in \{1,\dots,|A|-1\}$. 
Always guessing $X=0$ and $Z=0$ leads to equality. 
That this is optimal is to be expected, as it is easily seen that the state is a superposition of $X=0$ and $Z=0$ eigenstates. 
Thus the question of determining the region of allowed guessing probabilities, raised in \cite{renes_uncertainty_2016-1}, is completely solved. 

Moreover, \eqref{eq:tripartiteboundary} has an elegant geometric interpretation.
Letting $m=|A|$, $x=P_\text{opt}(X|B)_\psi$, and $z=P_\text{opt}(Z|C)_\xi$, it can be easily verified that the boundary is that of the ellipse  
\begin{align}
\label{eq:ellipse}
\frac{(x+z-1)^2}{1/m}+\frac{(x-z)^2}{(m-1)/m}= 1\,
\end{align}
in the region $1/m\leq x,z\leq 1$. 
For arbitrary $m$, these are precisely the ellipses that just fit in the unit square. 

Analogously to the use of \eqref{eq:maxbound} in the guessing game, \eqref{eq:maxboundq} implies a bound on monogamy of entanglement; specifically, on the possible values of $R(A|B)_{\rho,\cE}$ and $R(A|C)_{\rho,\cE'}$ for an arbitrary tripartite state $\rho_{ABC}$. 
In this case, using the duality of min and max entropy~\cite{konig_operational_2009}, which can be expressed as $\max_\sigma F(\rho_{AC},\pi_A\otimes \sigma_C)^2=R(A|B)_\rho$, we obtain 
\begin{align}
\label{eq:monogamyboundary}
|A|^2 R_\text{opt}(A|C)_\rho\leq \left( \sqrt{R_\text{opt}(A|B)_\rho}+\sqrt{|A|^2-1}\sqrt{1-R_\text{opt}(A|B)_\rho}\,  \right)^2\,.
\end{align}
Equality can be attained by a superposition of entanglement with $B$ and entanglement with $C$, namely $\ket{\Psi}_{ABC}=N^{-\nicefrac 12}\big(\cos\theta\ket\Phi_{AB}\ket 0_C+\sin\theta\ket\Phi_{AC}\ket 0_B\big)$, with the normalization constant $N=1+\sin2\theta/{d}$ for $d=|A|$.
Choosing trivial recovery maps, we obtain $R(A|B)_{\Psi,\cI}=(d \cos\theta+\sin\theta)^2/(d(d+\sin 2\theta))$ and $R(A|C)_{\Psi,\cI}=(d \sin\theta+\cos\theta)^2/(d(d+\sin 2\theta))$. 
Comparing \eqref{eq:monogamyboundary} and \eqref{eq:tripartiteboundary}, it is apparent that the latter satisfies the ellipse equation with $x=R(A|B)_{\Psi,\cI}$, $z=R(A|C)_{\Psi,\cI}$, and $m=d^2$. 
It is then straightforward to check that the particular values of $R(A|B)_{\Psi,\cI}$ and $R(A|C)_{\Psi,\cI}$ satisfy \eqref{eq:ellipse}. 

In fact, $\ket\Psi$ was used in \cite{kay_optimal_2009} to investigate the limits of what they term ``singlet monogamy'' and its relation to optimal cloning. 
The scenario they consider is nearly the same as here, except that the optimal channel in the recovery operation is restricted to be unitary and, importantly, they also consider monogamy involving more than two auxiliary parties. 
The state $\ket\Psi$ was conjectured to give the optimal bound, their Equation 6, and the above derivation shows that their conjecture holds true for monogamy relations of three systems.

\section{Discussion and open problems}
By simple application of the data processing inequality, we have given new and useful bounds involving fidelity and guessing probabilities of optimal and pretty good measurements, as well as for the corresponding quantities for entanglement recovery. 
These allow the complete characterization of the allowed guessing probababilities when two different parties try to simultaneously predict the value of one of two conjugate measurements on a quantum system, as well as an analogous statement for the allowed entanglement fidelities two parties can each locally create with a common system. 

It would be interesting to determine if the R\'enyi divergence at orders besides $\alpha=2,\infty$ is related to other particular guessing or entanglement recovery strategies, as this would immediately give new bounds. 
We can report the following partial result for $\alpha=3$ and the ``quadratically-weighted'' variant of the pretty good measurement, i.e.\ using $\Lambda_x=\bar\varphi^{-\nicefrac 12}p_x^2\varphi_x^2\bar\varphi^{-\nicefrac 12}$, for $\bar\varphi^{-\nicefrac 12}=\sum_x p_x^2\varphi_x^2$ (discussed, e.g.\ in \cite[Section 2.2]{belavkin_design_1987}). 
Suppose the $\varphi_x$ all commute, so $B$ is essentially a classical random variable $Y$. Then the average guessing probability in this case is $P_\text{quad}(X|Y)_\rho=\sum_y (\sum_x P_{XY}(x,y)^3)/(\sum_{x'} P_{XY}(x',y)^2)$. 
Bounding the denominator from above by $(\sum_{x'} P_{XY}(x',y))^2$, one finds that $D_3(\rho_{XB},\pi_X\otimes \rho_B)\leq \tfrac12\log |X|^2P_\text{quad}(X|B)$. 
Employing \eqref{eq:mainc} with $\alpha=3$ yields the relation
\begin{align}
P_\text{quad}(X|Y)_\rho\geq P_\text{opt}(X|Y)^3+\frac{(1-P_\text{opt}(X|Y))^3}{(|X|-1)^2}\,.
\end{align}
Unfortunately, this is weaker than the bound $P_\text{quad}(X|Y)_\rho\geq P_\text{opt}(X|Y)_\rho^2$ shown in \cite[Theorem 10]{tyson_two-sided_2009} and also valid for non-commuting ensembles. 
Nonetheless, this approach can presumably be easily extended to higher weights, e.g.\ cubic as considered in \cite{ballester_state_2008}, and may prove useful there.
One might also relate particular measurement strategies to other particular choices of the second argument to the divergence and investigate the implications of the data processing inequality in that context.
For instance, \cite[Theorem 4]{beigi_quantum_2014} shows that $\log |X| P_\text{pg}(X|B)_\rho\geq D_2(\rho_{XB},\tfrac1{|X|}\rho_{XB}+(1-\tfrac1{|X|})\rho_X\otimes \rho_B)$.

\vspace{.75\baselineskip}

\noindent{\bf Acknowledgments.} 
I thank Raban Iten for useful conversations. 
This work was supported by the Swiss National Science Foundation (SNSF) via the National Centre of Competence in Research ``QSIT''.

\printbibliography[heading=bibintoc,title=References]

\end{document}